\documentclass{aa}
\usepackage[varg]{txfonts}
\usepackage{lineno} 
\usepackage{amssymb,amsmath}
\usepackage{graphicx}
\usepackage{xcolor,natbib}
\usepackage{multirow}
\usepackage[T1]{fontenc}
\usepackage{ae,aecompl}
\usepackage{newtxtext, newtxmath}
\usepackage{float}
\usepackage{subfigure}
\usepackage{longtable}
\usepackage{enumitem}
\usepackage{longtable,listings}
\usepackage[flushleft]{threeparttable}
\usepackage{parcolumns}
\usepackage{hyperref}
\usepackage{CJK}
\usepackage{upgreek}
\usepackage{soul}

\def\oiii{[O~{\sc iii}]}

\def\obj{SDSS J1532}

\begin{document}

\title{SDSS J153231.80+420342.7: a triple black hole candidate with a close binary black hole}

\titlerunning{a triple black hole candidate}

\author{Qi Zheng \inst{\ref{inst}}
\and YiWen Jiang\inst{\ref{inst}}
\and Xue-Guang Zhang\inst{\ref{inst1}}
\and Qirong Yuan\inst{\ref{inst}}}

\institute{School of Physics and Technology, Nanjing Normal University, No. 1,	Wenyuan Road, Nanjing, 210023, P. R. China\label{inst} \\
\and Guangxi Key Laboratory for Relativistic Astrophysics, School of Physical Science and Technology, Guangxi University, Nanning, 530004, P. R. China\label{inst1} \\
\email{xgzhang@gxu.edu.cn; yuanqirong@njnu.edu.cn}}
\date{Received 08/12/2025 / Accepted 02/04/2026}

\abstract
{We report a triple black hole candidate with a close binary black hole (BBH) in the blue quasar SDSS J153231.80+420342.7 (=SDSS J1532) at a redshift of 0.209.
It shows double-peaked profiles in all narrow emission lines, which can be a signature of a dual AGN.
If the double-peaked features are produced by a dual AGN, the estimated physical separation between the two cores is about 3 kpc.
Alternative interpretations to the double-peaked profiles involving rotating disk-like narrow line region (NLR) and AGN-driven outflow models are also discussed for the double-peaked features.
Besides, SDSS J1532 shows optical quasi-periodic oscillations (QPO) of about 0.6 yr from the ZTF and CSS light curves, with time duration longer than 14 years, which may indicate a binary black hole with about 1 mpc separation.
Two alternative explanations, the disk precession and the jet precession models, are also discussed.
The current results cannot completely rule out alternative models for the characteristics of spectrum and light curves.
As a candidate for triple black hole with two cores in kpc scale and a close BBH in milli-pc scale in \obj, it may be going through a critical period in its evolution.}

\maketitle

\section{Introduction} 
Observational and theoretical studies have confirmed the presence of supermassive black holes (SMBHs) at the centers of most massive galaxies, with galaxy interactions being prevalent throughout the universe, leading to hierarchical evolution through collisions \citep{To72, Be80, Si98, Ho05, De23, Zh23}. Consequently, the presence of multiple SMBHs in massive galaxies should be common.
When the dual AGNs have projected separations of a few to several tens of kpcs, the two nuclei can be directly imaged \citep{Sh19,Zh23}. 
For more compact systems, double-peaked narrow emission lines (DPNELs), first introduced by \citet{Zh04} as indicators of dual AGNs at kpc scales, have been widely used in systematic searches \citep{Wa09, Sm10, Co12, Ge12, Li13, Ki20}. 

In the later stages of galactic mergers, when two nuclei approach each other at sub-pc distances, they form a BBH system, potentially giving rise to the signature of periodic electromagnetic radiation variations on month or year timescales due to the orbital motion of the two SMBHs.
There are many candidates for BBH systems determined through QPO in different bands as shown in \citet{Gr15,Se18,Se20,Mi22,Zxg23,Fo25,Hu25}.

In even rarer scenarios, triple black holes may emerge when a third galaxy subsequently merges before the coalescence of the initial black holes \citep{Va96}.
Up to now, most candidates of triple black hole systems show projected separations from a few to tens of kpcs \citep{Ba08,Li11,Ka17,Pf19,Fo21a,Fo21b,Pe22}, and only a few triple black hole candidates have the closest pairs separated by sub-kpc. 
\citet{De14} reported a triple black hole candidate, SDSS J150243.09+111557.3, with the closest pair separated by 140 pc and two components, which might be the reason for its double-peaked \oiii~emission lines, separated by 7.4 kpc.
However, the result of the closest pair was challenged by \citet{Wr14}, who thought that the radio components separated by 140 pc are double hot spots produced by a single SMBH.
\citet{Ko20} reported the presence of three distinct nuclei within a 1 kpc region in the late-stage merger system NGC 6240, and revealed that the previously unresolved southern component is in fact composed of two separate nuclei, with a projected separation of merely 198 pc.
Detecting pairs separated by so small distances through imaging is challenging due to angular resolution limitations.

One potential indicator of much closer distances with sub-pc relative separations is the detection of periodic electromagnetic radiation variations.
Therefore, it may also be a promising method for searching triple black hole systems by combining the DPNELs with QPO signatures.

In this paper, we investigate SDSS J153231.80+420342.7 (=SDSS J1532), a triple black hole candidate selected from a sample of type 1 AGNs with double-peaked \oiii~emission lines identified in the Data Release 16 of the Sloan Digital Sky Survey \citep{Zh25}. The virial black hole mass, estimated from the broad H$\alpha$ emission line, is $10^{8.68\pm 0.02}\rm M_\odot$.
In addition to the prominent double-peaked narrow emission-line profiles, SDSS J1532 shows evidence for QPO in the optical band.
These properties suggest that SDSS J1532 may have two compact cores separated by sub-pc distances and a third core located at a kpc-scale distance away from them.
The data and methods used for SDSS J1532 are shown in Section 2, the results and discussions of possible physical models are presented in Section 3, and the summary and conclusions are given in Section 4.
We have adopted the cosmological parameters of $H_{0}=70 {\rm km/s/Mpc}$, $\Omega_{\Lambda}=0.7$ and $\Omega_{\rm m}=0.3$.

\section{Data and Methods}

\subsection{Spectrum}
\begin{figure*}
	\centering\includegraphics[width=18cm,height=12cm]{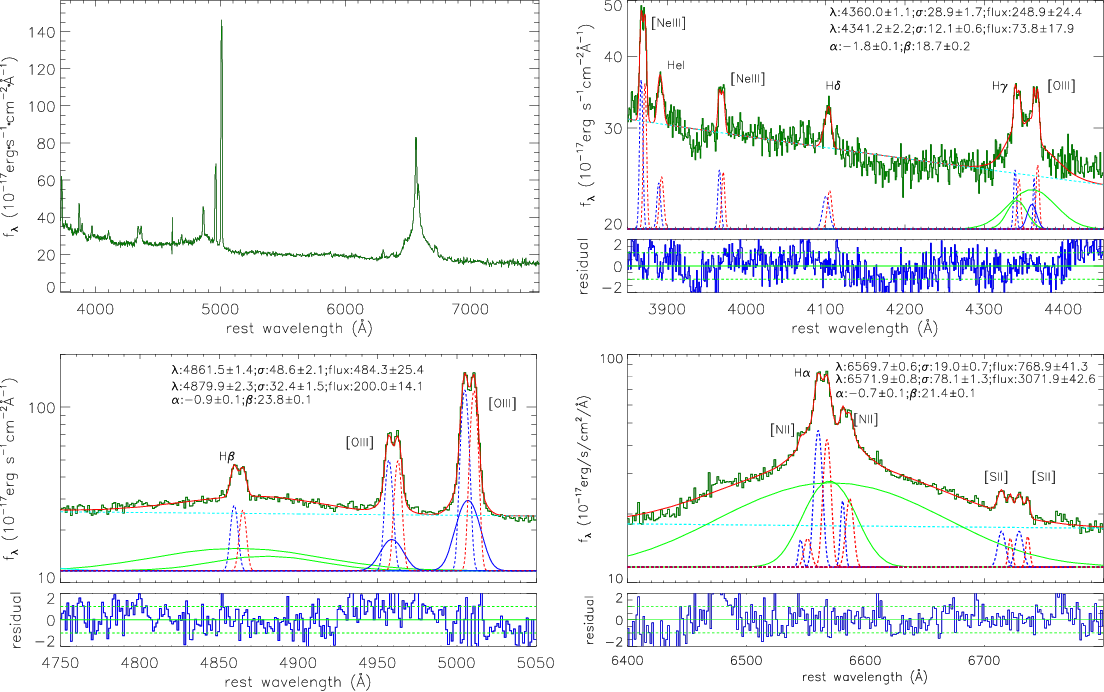}
	\caption{The rest-frame optical spectrum of SDSS J1532 (top-left panel), and best fitting results of SDSS J1532 with rest-frame spectrum from 3850\AA~to 4450\AA~(top-right panel), from 4750\AA~to 5050\AA~(bottom-left panel), and from 6400\AA~to 6800\AA~(bottom-right panel).	
		In the top of the panels, the solid dark green lines show the line spectra; the dashed blue and red lines represent the blue-shifted and red-shifted components of narrow H$\alpha$, H$\beta$, H$\gamma$, H$\delta$, He~{\sc~i}, [Ne{\sc~iii}], \oiii, [N{\sc~ii}] and [S{\sc~ii}] doublets; the solid blue lines represent the extended components of \oiii~emission lines; the solid green lines represent broad H$\alpha$, H$\beta$ and H$\gamma$ emission lines with corresponding parameters of central wavelength (\AA), line width $\sigma$ (\AA) and line flux ($10^{-17}{\rm erg/s/cm^{2}}$) listed in the first two lines of the panels; the dashed cyan lines show the continuum emissions $\alpha\times(\frac{\lambda}{5100\textsc{\AA}})^{\beta}$ with parameters listed in the third line.
		In the bottom of the panels, the solid blue lines represent the residuals computed by subtracting the best-fitting results from the line spectra and then dividing by the uncertainties of the SDSS spectra; the solid and dashed green lines show 0 and $\pm$1, respectively.
	}
	\label{fig1}
\end{figure*}

SDSS J1532 is a typical QSO with broad emission lines at z $\sim$0.209.
The processed and calibrated spectrum \footnote{\url{https://www.sdss.org/dr14/spectro/pipeline/}} is downloaded from the Data Release 16 of the Sloan Digital Sky Survey (SDSS) \citep{Ah20}, and it is presented in the top-left panel of Figure \ref{fig1}.
This SDSS spectrum, obtained on MJD=53149, is taken with a 2.5-m wide-field telescope using a fiber diameter of $3^{\prime\prime}$, and covers the wavelength range 3794–9189 \AA~sampled by 3843 pixels. The spectral resolution is R$\approx$1800.

We focus on the main emission lines, spanning the rest wavelength from 3850 \AA~to 4450 \AA, from 4600\AA~to 5050\AA~and from 6250 \AA~to 6900 \AA.
The following model functions are used to describe these emission lines.
A single power law component is applied to describe continuum emission underneath the H$\gamma$, H$\beta$, H$\alpha$ emission lines, respectively. Two narrow Gaussian functions (the second moment $\sigma$<400 km/s) are applied to describe the double-peaked profiles in each narrow emission line, and
an additional Gaussian component ($\sigma$>400 km/s) is included to model the wing component of \oiii.
Finally, two broad Gaussian functions ($\sigma$>800 km/s) are used to model the broad components of the H$\gamma$, H$\beta$, H$\alpha$ lines.
Here, the positions of the emission lines are determined relative to their laboratory rest wavelengths \citep{Va01}.

The parameters are then constrained as follows: 
For the \oiii, [N~{\sc ii}], [Ne~{\sc iii}] doublets, the central wavelengths and line widths of the blue-shifted and red-shifted components are fixed with those of their respective companion lines within each doublet, and the flux ratio set to the theoretical value of 3:1 \citep{St00}. 
For the [S~{\sc ii}] doublet, the line width of the blue-shifted and red-shifted components are consistent with those of their respective companion lines in velocity space.
The central wavelength of blue-shifted (red-shifted)  component of the Balmer narrow emission lines is fixed together in velocity space.

The main emission lines of SDSS J1532 are fitted using the Levenberg–Marquardt least-squares minimization method, as implemented in the MPFIT package \citep{Ma09}. The results ($\chi^2$/$dof$=1.3, $dof$=1373) are shown in Table 1.
Furthermore, the peak separations and their associated uncertainties are derived from the central wavelengths and their corresponding uncertainties, and are presented in Table 1.
To clearly display the double-peaked profiles, portions of the rest-frame spectrum spanning 3850\AA~- 4450\AA, 4750\AA~- 5050\AA, and 6400\AA~- 6800\AA~are shown in Figure \ref{fig1}, with the best-fitting model overplotted.

\setlength{\tabcolsep}{4pt}
\begin{table}
\caption{Features of main narrow emission lines}
\begin{center}
\begin{tabular}{ccccc}
\hline\hline
Line& $\lambda$ & $\sigma$  &  Flux  & $\Delta\upsilon$ \\ \hline
\multirow{2}{*}{[Ne \sc{iii}]}	& 3867.1$\pm$ 0.3   & 1.9$\pm$0.2 & 80.1$\pm$4.9 & \multirow{2}{*}{363.5$\pm$50.3}  \\ 
{ }	& 3871.8$\pm$0.3   & 1.9$\pm$0.1 & 76.7$\pm$4.8 & { }  \\ 
\hline
\multirow{2}{*}{He \sc{i}}	& 3889.2$\pm$ 0.5   & 2.3$\pm$0.3 & 23.3$\pm$14.1 & \multirow{2}{*}{277.7$\pm$123.4}  \\ 
{ }	& 3892.8$\pm$1.1   & 2.6$\pm$0.4 & 30.3$\pm$14.2 & { }  \\ 
\hline
\multirow{2}{*}{H$\delta$}	& 4100.2   & 4.1$\pm$0.5 & 29.0$\pm$4.4 & \multirow{2}{*}{339.3}  \\ 
{ }	& 4104.8   & 2.7$\pm$0.7 & 22.3$\pm$6.7 & { }  \\ 
\hline
\multirow{2}{*}{H$\gamma$}	& 4338.8   & 1.8$\pm$0.3 & 25.8$\pm$5.6 & \multirow{2}{*}{339.3}  \\ 
{ }	& 4343.7   & 2.1$\pm$0.6 & 22.8$\pm$7.7 & { }  \\ 
\hline
\multirow{2}{*}{[O \sc{iii}]}	& 4362.7$\pm$0.5   & 1.2$\pm$0.5 & 13.7$\pm$6.4 & \multirow{2}{*}{323.1$\pm$68.7}  \\ 
{ }	& 4367.4$\pm$0.5   & 2.0$\pm$0.5 & 28.5$\pm$6.5 & { }  \\ 
\hline
\multirow{2}{*}{H$\beta$}	& 4859.5$\pm$0.1   & 2.1$\pm$0.2 & 83.4$\pm$8.3 & \multirow{2}{*}{339.3$\pm$12.3}  \\ 
{ }	& 4865.0$\pm$0.2  & 2.0$\pm$0.2 & 70.4$\pm$8.2 & { }  \\ 
\hline
\multirow{2}{*}{[O \sc{iii}]}	&5005.0$\pm$0.1  & 2.1$\pm$0.1 & 604.5$\pm$17.8  & \multirow{2}{*}{347.4$\pm$12.0} \\ 
{ }	& 5010.8$\pm$0.1  & 2.0$\pm$0.1 & 576.7$\pm$17.4  & {  } \\ 
\hline	
\multirow{2}{*}{H$\alpha$}	& 6560.3  & 2.9$\pm$0.2 & 250.7$\pm$10.8 & \multirow{2}{*}{339.3} \\ 
{ }	& 6567.7  & 2.8$\pm$0.2 & 214.5$\pm$10.5 & { } \\ 
\hline	
\multirow{2}{*}{[N \sc{ii}]} 	& 6580.8$\pm$0.4 & 1.7$\pm$0.3 & 47.1$\pm$5.9 & \multirow{2}{*}{273.3$\pm$45.6} \\ 
{ } 	& 6586.8$\pm$0.6 & 2.9$\pm$0.5 & 84.6$\pm$8.7 & { } \\ 
\hline
\multirow{2}{*}{[S \sc{ii}]} 	& 6714.5$\pm$0.5 & 3.4$\pm$0.5 & 45.8$\pm$5.6& \multirow{2}{*}{326.0$\pm$40.2} \\
{ } 	& 6721.8$\pm$0.4  & 1.5$\pm$0.3 & 14.6$\pm$5.4 & { } \\
\hline
\multirow{2}{*}{[S \sc{ii}]} 	& 6729.2$\pm$0.5  & 3.4 & 42.7$\pm$6.1 & \multirow{2}{*}{329.7$\pm$40.1} \\
{ }	& 6736.6$\pm$0.4  & 1.5 & 15.8$\pm$ 4.0 & { } \\ \hline
\hline
\end{tabular}
\begin{tablenotes}
\item
The units of the central wavelength $\lambda$ and line width $\sigma$ are \AA, the units of the line flux are $10^{-17}{\rm erg/s/cm^{2}}$, and the units of the peak separation $\Delta\upsilon$ are km/s.
\end{tablenotes}
\end{center}	
\end{table}

\subsection{Optical light curves}
The Zwicky Transient Facility (ZTF; 47 deg$^2$ field of view) \citep{Be19,Gr19,De20}, 
dedicated to surveying the transient and variable universe,
operates on the 48-inch Samuel Oschin telescope at Palomar Observatory with three custom-made filters: ZTF-g, ZTF-r, and ZTF-i.
The Catalina Sky Survey (CSS) \citep{Dr09}, supported by the Near Earth Object Observation Program, is dedicated to the discovery and track of near-Earth objects. This survey is analyzed by the Catalina Real-Time Transient Survey to detect optical transient phenomena (V < 21.5 mag).

\begin{figure*}
\centering\includegraphics[width=8cm,height=5cm]{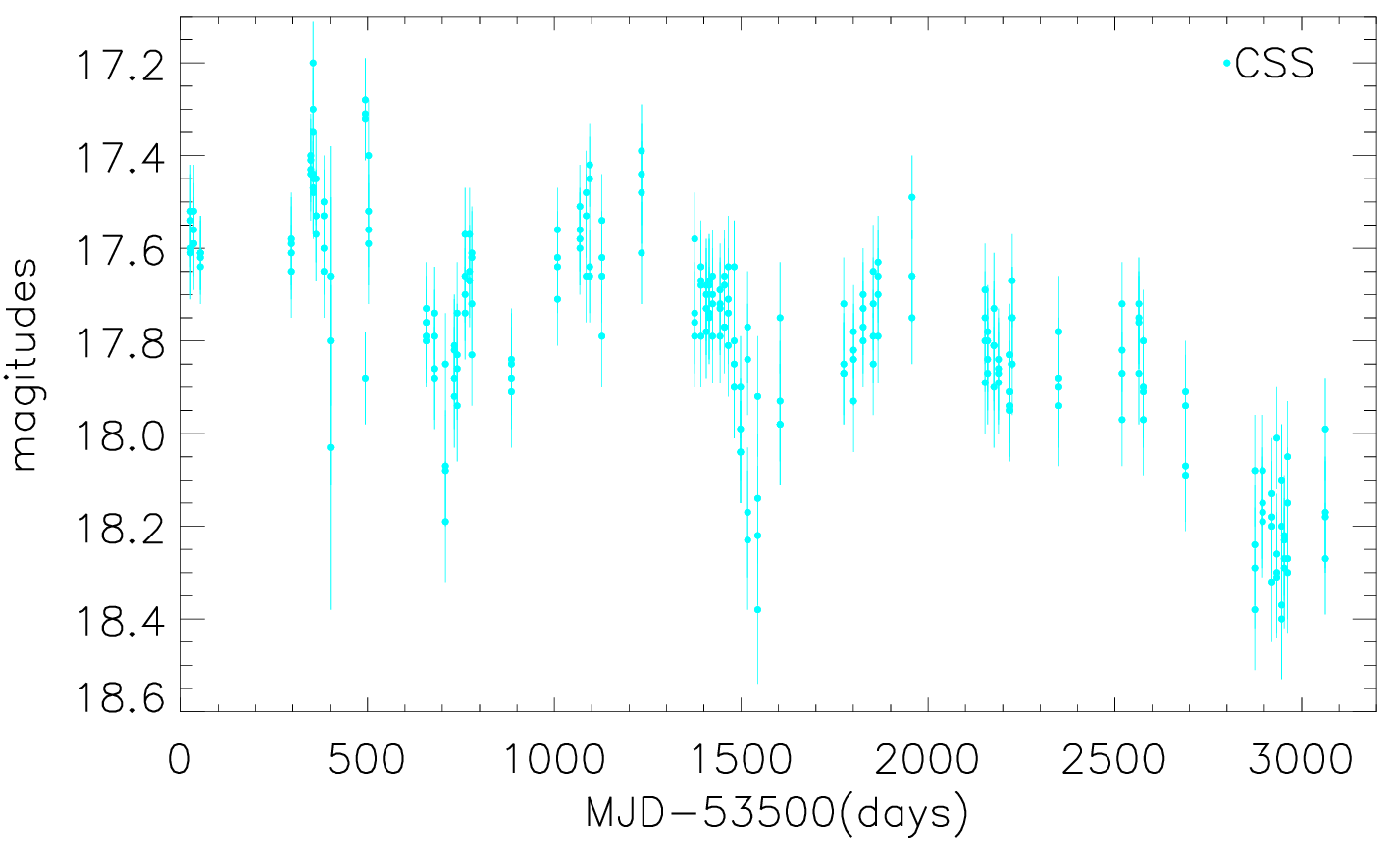}
\centering\includegraphics[width=8cm,height=5cm]{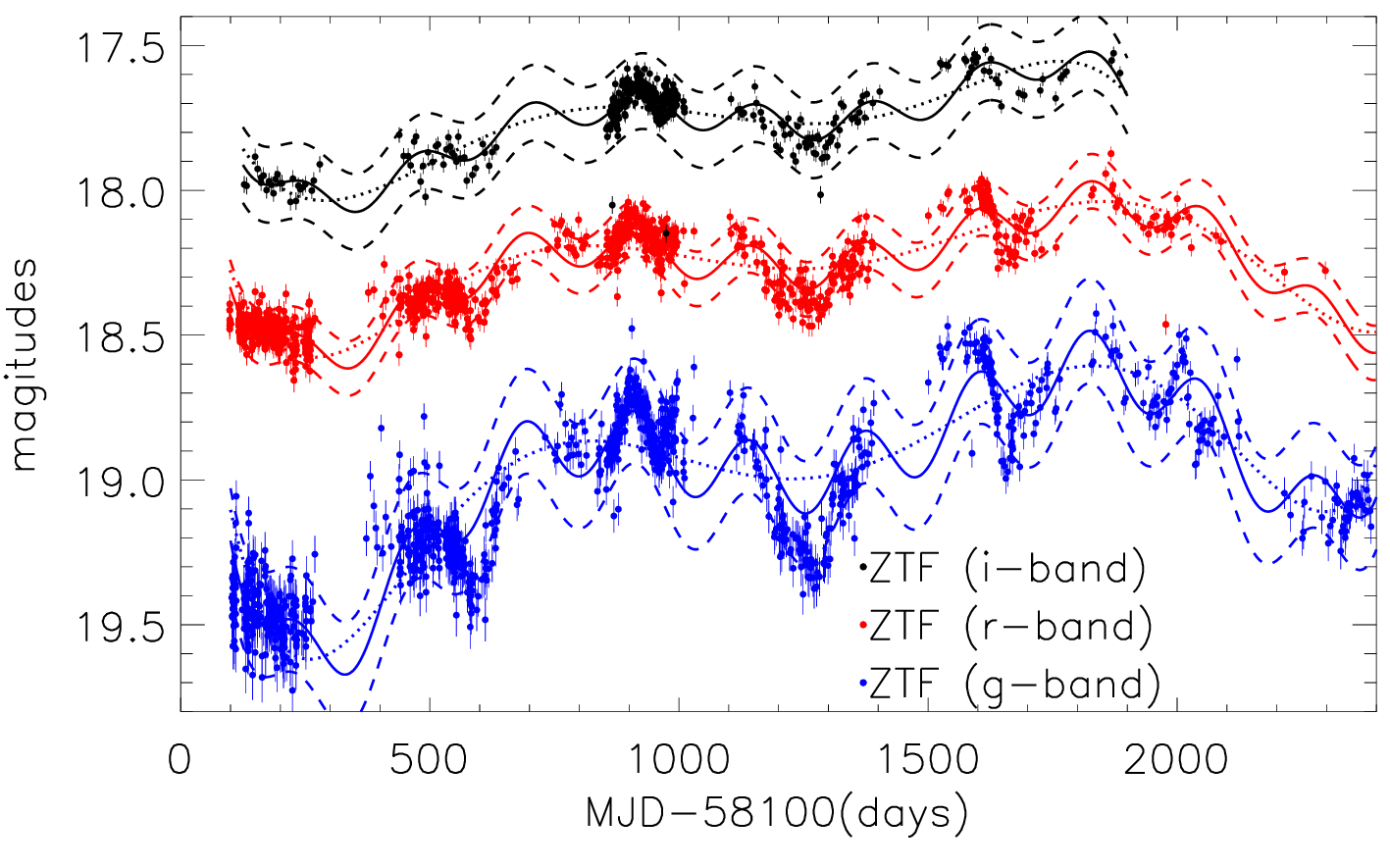}
\caption{The observed optical light curves of SDSS J1532 directly downloaded from the survey archives of the CSS (v-band) and ZTF (g-band, r-band, i-band). The solid and dashed lines show the best-fitting results and their corresponding 1RMS scatter bands, respectively. The dotted lines represent the results of polynomial function.}
\label{fig2}
\end{figure*}

The optical-band photometric data of SDSS J1532 from ZTF (MJD: 58202-60489 for the g-band; 58198-60397 for the r-band; 58227-59985 for the i-band)
and CSS (MJD: 53552-56563) are shown in Figure \ref{fig2}.
Due to lower quality of the CSS light curve, we only apply the epoch-folding method discussed later, without conducting further analysis of its periodicity.

A simple sinusoidal function plus a six-degree polynomial function ($a+bt+ct^2+dt^3+et^4+ft^5+gt^6+hsin(2\pi t/T+\phi)$) is applied to describe the three band light curves in ZTF with the same periodicity $T$, respectively.
Here, the sine component is only applied to show the periodic variability patterns related to QPOs, without delving into the physical origin of the QPOs, while the polynominal component is only applied to trace the long-term non-periodic trends.
Such trends could be commonly attributed to stochastic accretion-rate fluctuations, thermal or viscous instabilities in the accretion disk \citep{Ke09,Ma10,De11}.
The best fitting results, leading to $T=226\pm9$ days and $\chi_1^2/dof_1=4.55$ ($dof_1 = 2690$), are obtained by the Levenberg–Marquardt least-squares minimization technique, as shown in the right panel of Figure \ref{fig2}.
The necessity of the polynomial term is verified statistically. A model including only a linear trend plus a sinusoid yields $\chi_2^2/dof_2 = 10.00$ with $dof_2 = 2705$ ($T = 219 \pm 1$ days). The inferred periodicity is robust against the choice of detrending model. An F-test \citep{Ma97,Ge12,Zh25} indicates that this improvement is significant with $>5\sigma$ confidence level, demonstrating that the higher-order polynomial is required to remove slow long-term variations.

To provide further evidence supporting the sinusoidal component, only six-degree polynomial function is applied to re-describe the light curves, resulting in  $\chi_3^2/dof_3=6.46$ ($dof_3 = 2697$).
Utilizing the F-test technique, we conclusively determine, with a confidence level exceeding 5$\sigma$, that the determined sinusoidal component plus polynomial function is preferred.

To further explore and robustly confirm the QPO properties of SDSS J1532, three other methods are adopted: the Lomb–Scargle periodogram (LSP) method, the weighted wavelet Z-transform (WWZ) method \citep{Fo96,Te04,Sa20}, and the epoch-folding method. 
Given that the r-band and g-band data from ZTF provides more data points and a longer observation period, the original data from these two bands are used to analyze periodicity.

\begin{figure}
	\centering\includegraphics[width=8cm,height=5cm]{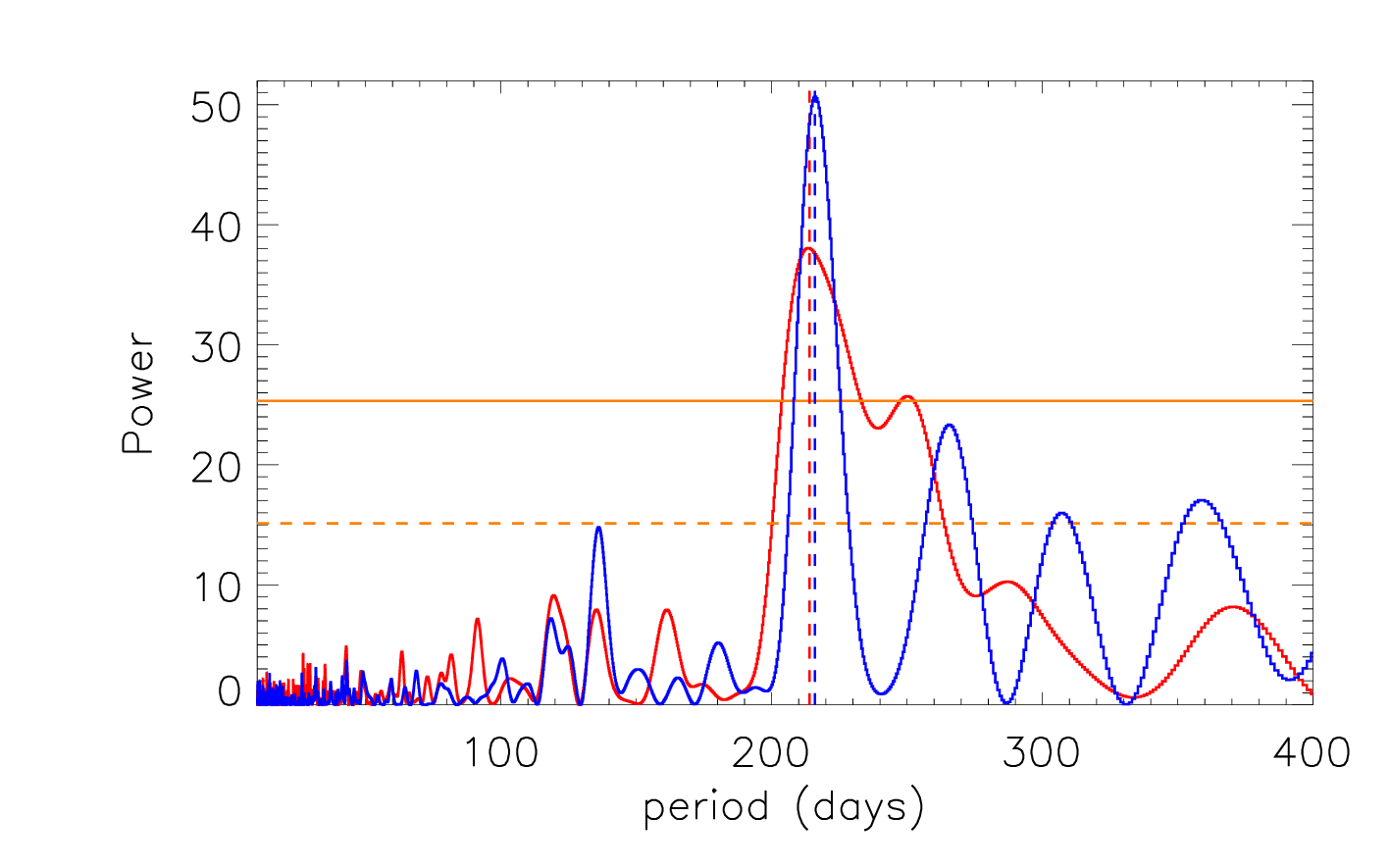}
	\caption{The results of LSP method in ZTF r-band (solid red line) and g-band (solid blue line) light curves.
		The vertical red line and blue line mark the position of the corresponding peaks of the power, the dashed and solid orange lines show 3$\sigma$ and 5$\sigma$ significance level determined by false alarm probability, respectively. 
	}
	\label{fig3}
\end{figure}

The classical Lomb–Scargle periodogram is widely employed for efficient periodicity search, utilizing least-square fitting of sinusoidal waves to accommodate data gaps and irregularities \citep{Ba23}.
Here, the IDL package $\sc{scargle}$ \footnote{\url{https://github.com/emrahk/IDL\_General/blob/master/third\_party/aitlib/timing/scargle.pro}} is applied with 10000 independent frequencies in frequency range from 0.0025 to 0.1 day$^{-1}$, and the results are shown in Figure \ref{fig3}. The lower frequency here corresponds to about $1/2T$, where $T$ is the periodicity derived from Figure~\ref{fig2}. 
For the Lomb–Scargle periodogram analysis, the observed light curves in each band are first detrended by subtracting a linear trend to remove long-term trends, and the resulting residual light curves are used as the input data.
The ZTF r-band light curve reveals a periodicity of 214$\pm$6 days and ZTF g-band light curve reveals a periodicity of 216$\pm$4 days with a significance level exceeding 5$\sigma$ determined by false alarm probability. 
The uncertainty of the periodicity is determined using the widely applied bootstrap method as follows. 
One third of the data points from the observed ZTF r-band (g-band) light curve are randomly selected to reconstruct a new light curve. From the 1000 reconstructed light curves, new periodicities are determined using the same LSP power properties. The uncertainties are assessed based on the half-width at half-maximum of the Gaussian-like distributions of these new periodicities.

\begin{figure*}
	\centering\includegraphics[width=8cm,height=5cm]{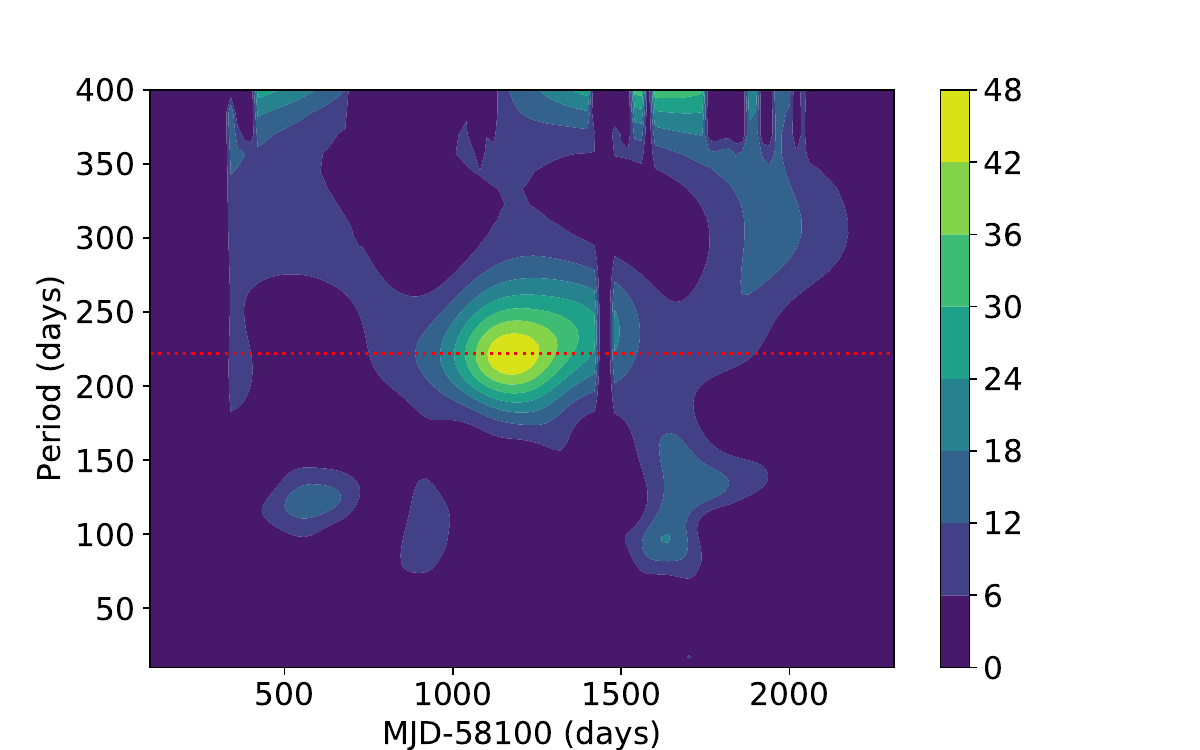}
	\centering\includegraphics[width=8cm,height=5cm]{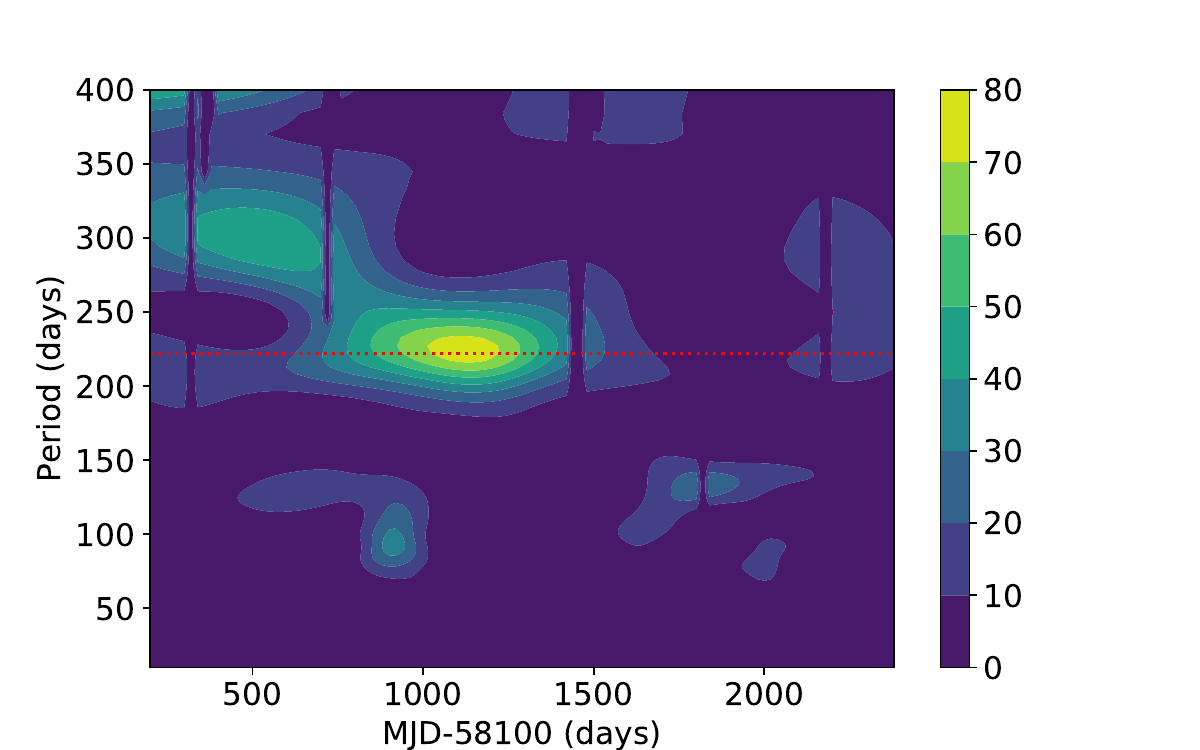}
	\caption{The results of WWZ method in ZTF r-band (left panel)
		and g-band (right panel) light curves of SDSS J1532.
		In each panel, the horizontal red line represents the position of the corresponding periodicity.
	}
	\label{fig4}
\end{figure*}

\begin{figure*}
	\centering\includegraphics[width=5.9cm,height=3.7cm]{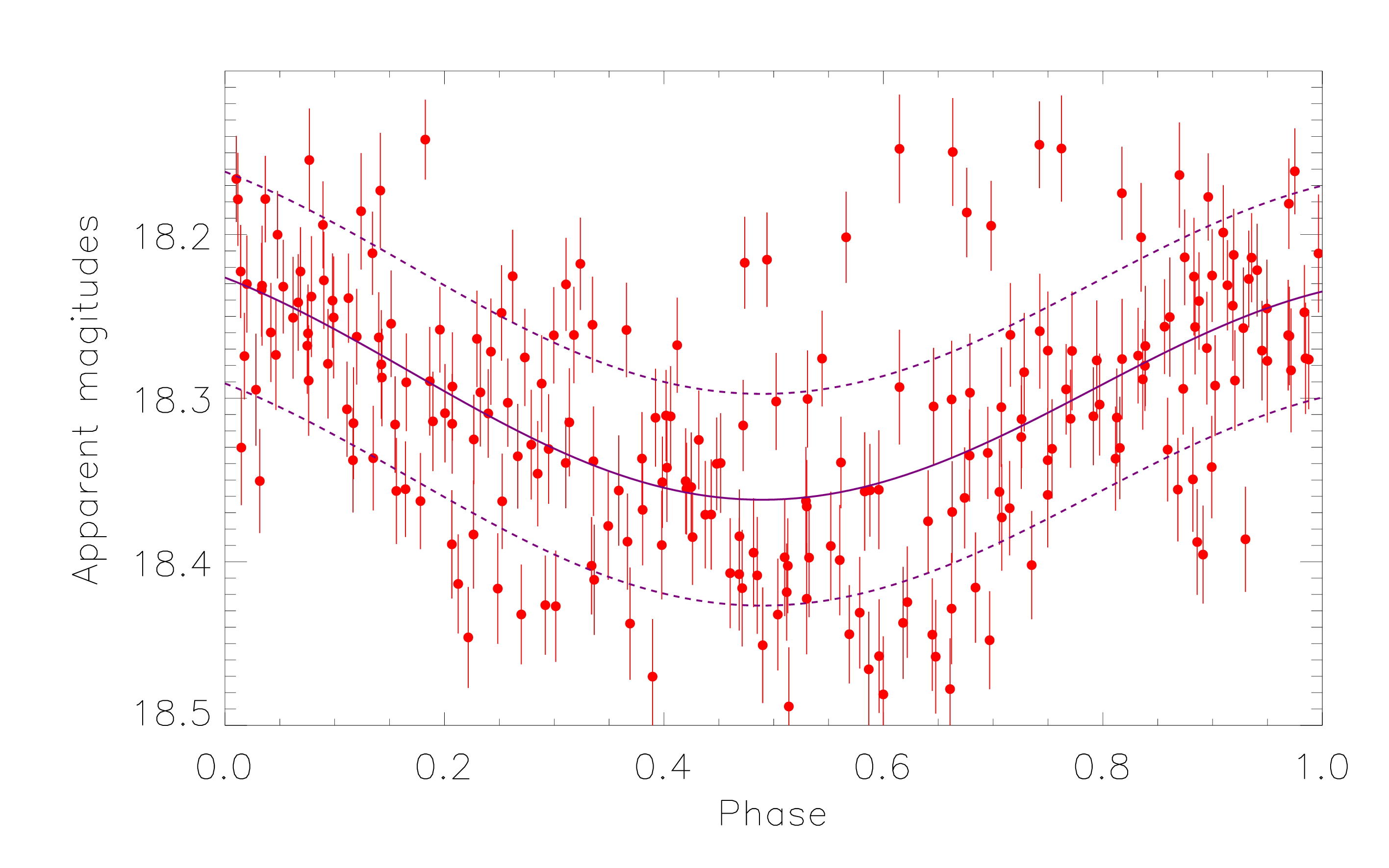}
	\centering\includegraphics[width=5.9cm,height=3.7cm]{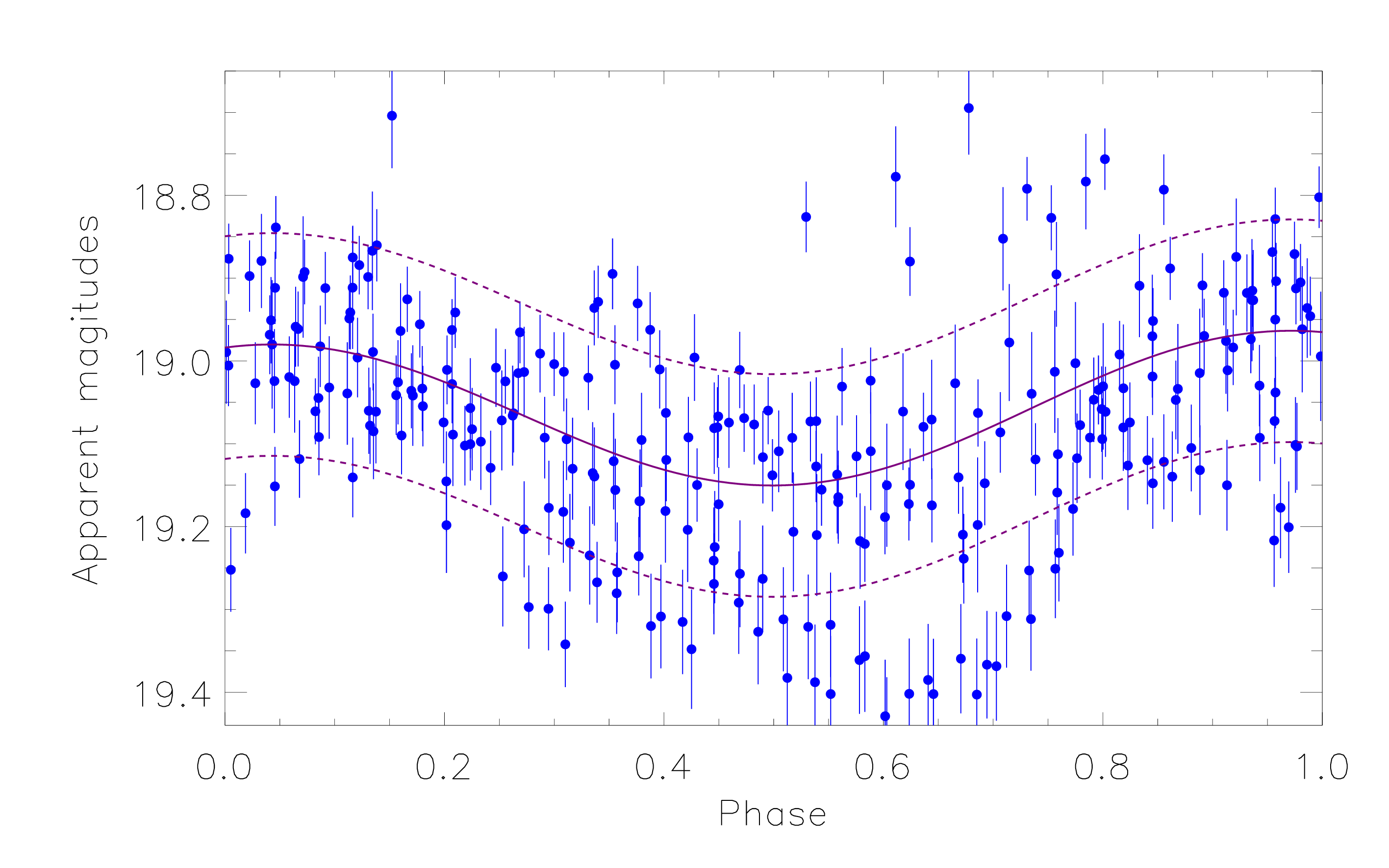}
	\centering\includegraphics[width=5.9cm,height=3.7cm]{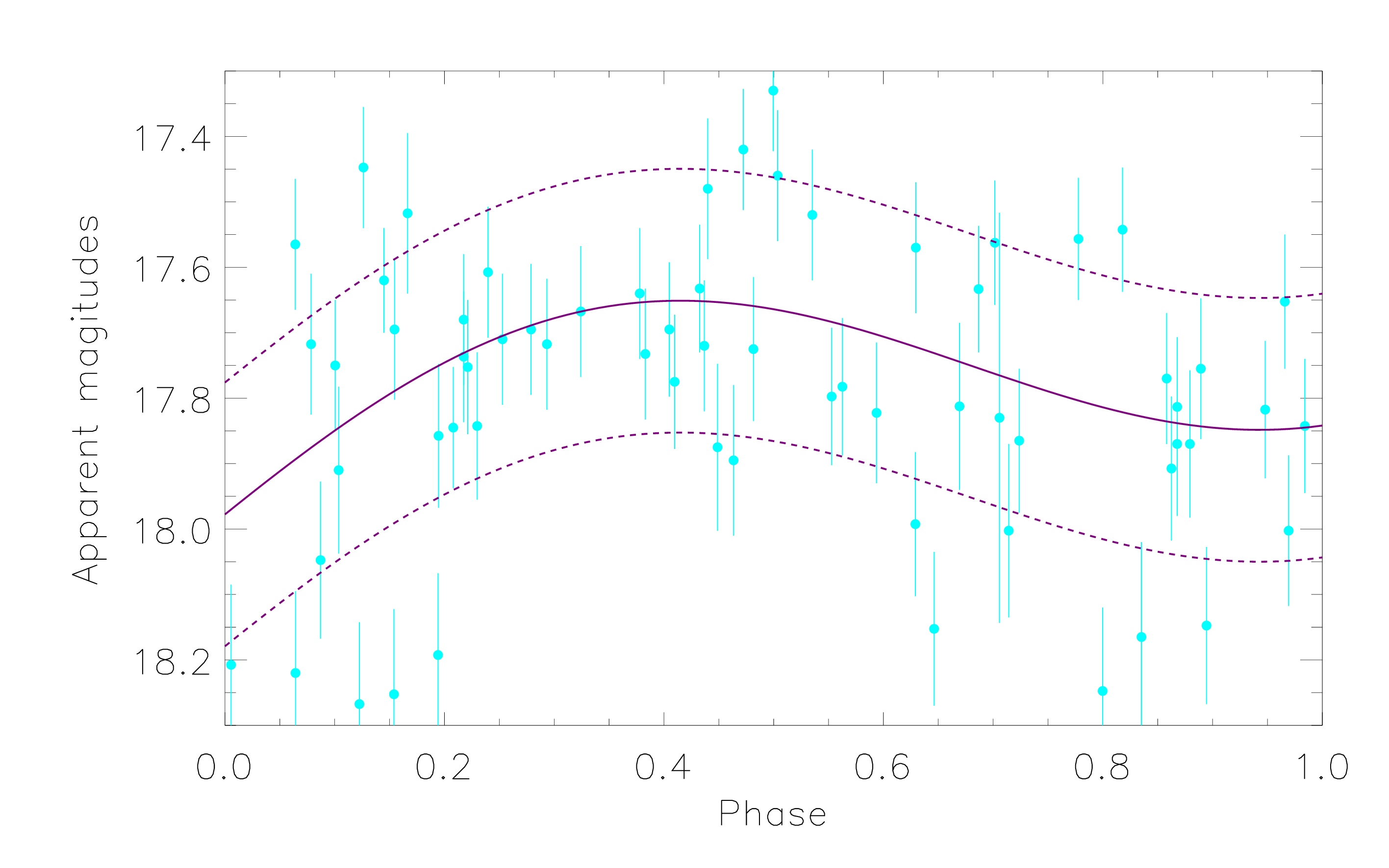}
	\caption{The epoch-folding method results of ZTF 5-day binned r-band light curve with a 226 day periodicity (left panel),
		ZTF 5-day binned g-band light curve with a 225 day periodicity (middle panel) after subtracting the contributions of the six-degree polynomial function, and CSS 1-day binned light curve with a 223 day periodicity (right panel).
		In each panel, the solid and dashed purple lines show the best-fitting descriptions by sinusoidal function and their corresponding 1RMS scatters, respectively.
	}
	\label{fig6}
\end{figure*}

The WWZ method, based on wavelet analysis and vector projection, employs z-statistics for period extraction. It enhances the performance of traditional wavelet techniques for realistic data, particularly when the data are sparse and unevenly sampled. We employed the Python code from \footnote{\url{https://github.com/eaydin/WWZ}} to calculate the WWZ power with frequency step of 0.0001 in frequency range from 0.0025 to 0.1 day$^{-1}$, and the results are illustrated in  Figure \ref{fig4}.
The periodicities of 221$\pm$15 days in ZTF r-band and 220$\pm$6 days in ZTF g-band light curves are identified by the WWZ method, which are consistent with those obtained from the LSP method, and the uncertainties are determined by bootstrap method as described above.

The presence of red noise in the optical band can significantly influence the detection of QPOs. . 
Red noise not only has the potential to obscure genuine QPO signals, but can also mimic periodic variability through a few-cycle, sinusoid-like modulation \citep{Va16}.
Therefore, it is important to evaluate the likelihood of spurious QPO detections in light curves generated purely from stochastic processes such as the continuous autoregressive (CAR) / damped random walk (DRW) models with the basic parameters of the intrinsic variability amplitude $\sigma$ and the intrinsic variability timescale $\tau$.
A comparison between a pure red-noise (DRW) model and a DRW model with an additional sinusoidal component may not always provide a clear distinction regarding the presence of periodic variability. In practice, the DRW model alone can reproduce a sinusoidal-like signal reasonably well under realistic sampling conditions, suggesting that stochastic red noise may partially mimic or absorb a periodic component, while QPO signals in AGNs are generally not perfectly sinusoidal. In addition, \citet{Wi22} showed that periodic signals become increasingly difficult to distinguish from red noise when the signal amplitude is small or comparable to the stochastic variability. Therefore, in this work we do not adopt a DRW + sinusoidal model to describe the light curve of SDSS J1532.

To this end, we generate 100,000 artificial light curves using the CAR process with the following parameters: a characteristic timescale of $\tau$ = 188 days, as determined by the publicly available JAVELIN code \citep{Ko10,Zu13} in the ZTF g-band light curve, and a long-term variance $\tau\sigma^2/2$ = 0.085, consistent with the observed variance of the ZTF g-band light curve. 
Given the larger number of data points in the g-band, the synthetic light curves generated in this work adopt the same time sampling as the original ZTF g-band data.
We then apply the following three criteria to identify potential false QPO detections among the simulated light curves: (1) The light curve must exhibit a strong periodic signal, with a LSP power greater than 5$\sigma$ significance level and a detected period between 180 and 260 days. 
(2) The light curve must be well described by a model consisting of a sinusoidal function added to a sixth-degree polynomial trend, with a $\chi^2/\rm dof$ less than 8 (two times larger than that of the observed light curve) and a fitted period within the same range (180–260 days).
(3) The sinusoidal amplitude in the fitted artificial light curves is greater than 0.06 after considering that the sinusoidal fit to the ZTF g-band light curve yields an amplitude of 0.12.
Among the 100,000 CAR-generated light curves, there are 178
light curves satisfying all three criteria above.
So the probability of detecting fake QPOs is 0.178\% (corresponding to higher than 3$\sigma$ confidence level).

To assess the reliability of the periodicities determined by the WWZ and LSP methods, we employ the epoch-folding method
to describe the light curves through a sinusoidal function. 
After subtracting the contributions of the six-degree polynomial function, the ZTF r-band and g-band light curves are folded, and the best fitting periods are 226 ($\chi^2/dof=4.89$,$dof$=236) and 225 days ($\chi^2/dof=7.07$,$dof$=283) for the ZTF r-band and g-band, respectively. 
Here, the subtractions of the polynomial components can lead to more apparent folded results.
It can be accepted that there is little difference in the best fitting periods between the ZTF r-band and g-band light curves because the light curves in different bands are fitted independently and have slightly different sampling and noise properties.
The best fitting periods for CSS light curve is 223 days ($\chi^2/dof=3.54$,$dof$=61).
Figure \ref{fig6} shows best fitting results by the epoch-folding method. The relatively large $\chi^2/dof$ values reflect the presence of stochastic, red-noise–like variability that is not captured by a simple sinusoidal model; therefore, the fits are used only to characterize the periodic modulation rather than to fully reproduce the detailed light curve structure.
Moreover, each of the three folded light curves is also described by a one-degree polynomial, and the $\chi^2/dof$ determined by the best fitting results are 6.88, 8.51, and 4.63, respectively.
Based on the F-test technique, a sinusoidal function is preferred to a one-degree polynomial with confidence level higher than 5$\sigma$.

\begin{figure*}
	\centering\includegraphics[width=8cm,height=5cm]{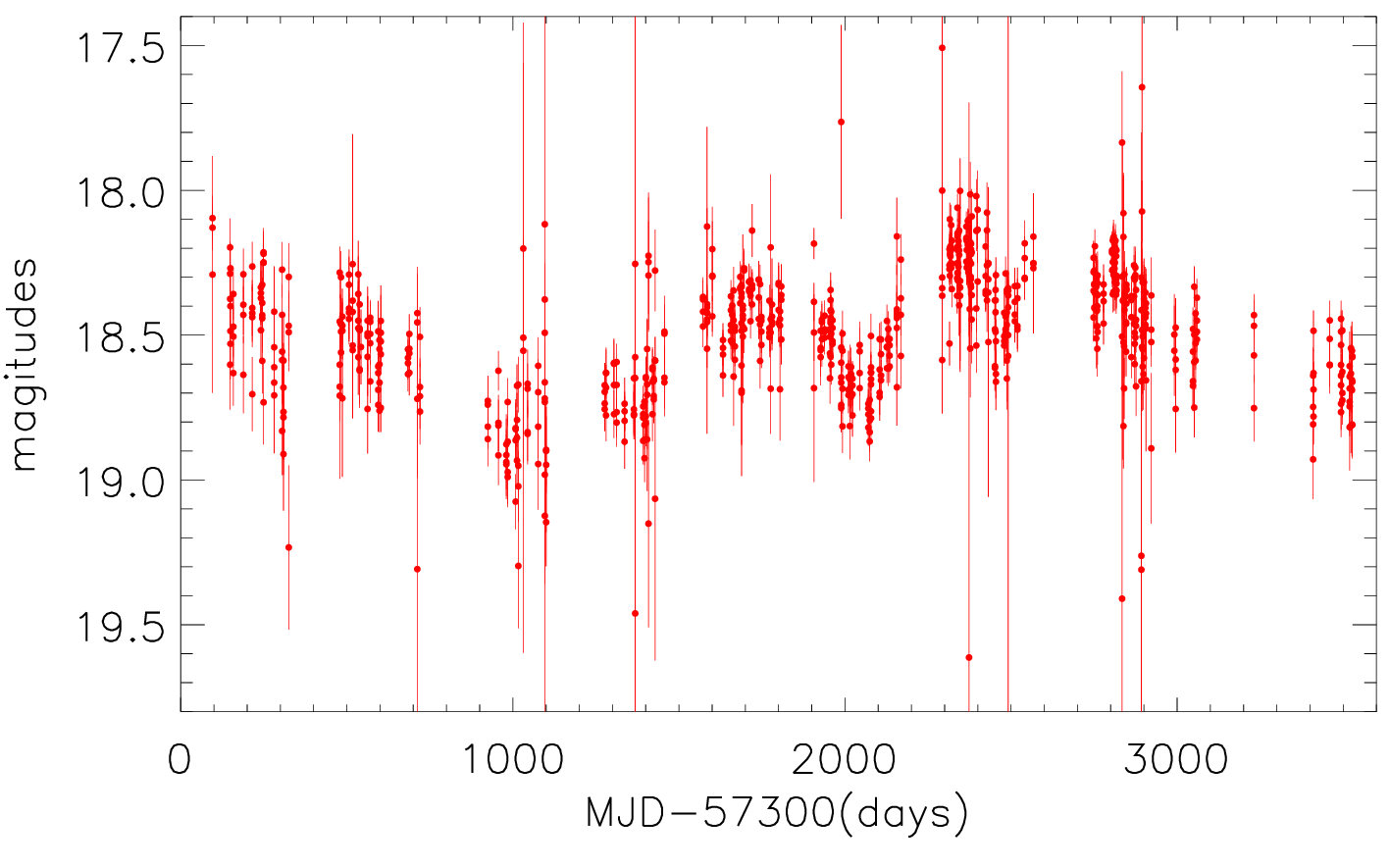}
	\centering\includegraphics[width=8cm,height=5cm]{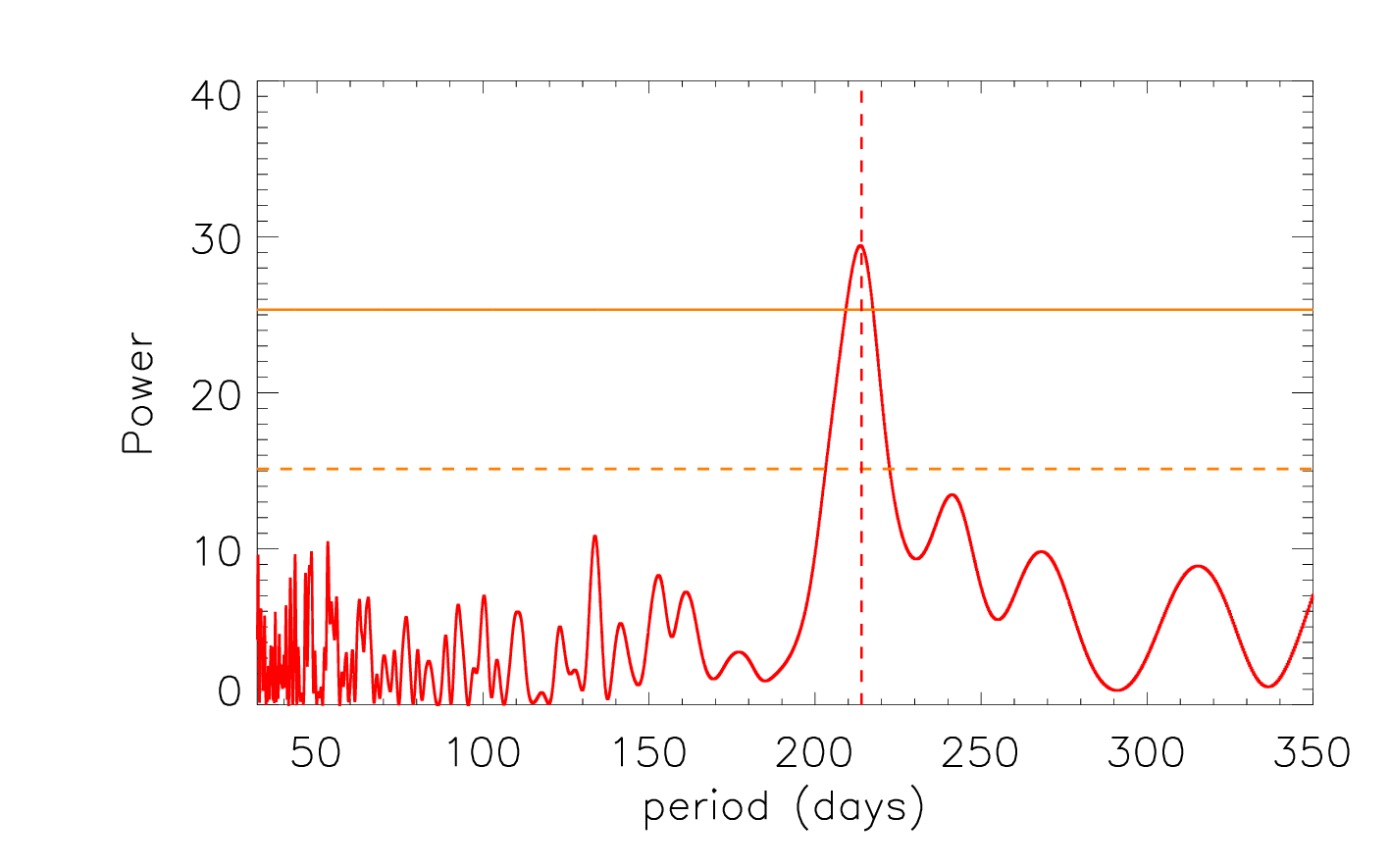}
	\caption{The ATLAS c-band light curve of SDSS J1532 (left panel) and the result of LSP method in the ATLAS c-band light curve (right panel). In the right panel, the vertical red line mark the position of the peak of the power, the dashed and solid orange lines show 3$\sigma$ and 5$\sigma$ confidence level determined by false alarm probability, respectively. 
	}
	\label{atlas}
\end{figure*}

Moreover, as discussed in \citet{Va18}, the various observational patterns inevitably imprint on the power spectrum derived from the observed data. 
To confirm the reliability of the observed peak, we search the Asteroid Terrestrial-impact Last Alert System (ATLAS; \citealt{To18}) c-band data of SDSS J1532, and present its light curve in the left panel of Figure \ref{atlas}. Given that the ATLAS light curve covers a similar time span to that of the ZTF but has larger uncertainties, we apply the LSP method to the ATLAS c-band light curve solely for the purpose of testing periodicity, in order to eliminate the possibility of spurious signals arising from various observational patterns. As shown in the right panel of Figure \ref{atlas}, we find that the ATLAS c-band also exhibits a periodicity of approximately 214 days.
The data used in this analysis are also obtained by subtracting a first-order polynomial function from the observed light curves.
Thus, the periodicity of about 220 days for SDSS J1532 is not the result of the observational pattern.

\section{Main Results and Discussions}

\subsection{Spectroscopic features}
As discussed in \citet{Xu09,Zh24}, objects with double-peaked profiles in all narrow emission lines could strengthen the presence of kpc-scale dual AGN candidates.
However, most objects exhibit double-peaked profiles only in \oiii~\citep{Zh04,Ki20,Zh25}, while those with prominent double peaks across all narrow emission lines are predominantly found in type 2 AGNs \citep{Wa09,Xu09,Se21}. 
Only a small number of type 1 AGNs display such profiles consistently in all narrow lines \citep{Ge12,Zh24}.
SDSS J1532 is a good candidate for a kpc-scale dual AGN with double-peaked profiles in all of its narrow emission lines.
In addition, NLR kinematics, such as
rotating disk-like NLR and AGN-driven outflow \citep{Cr10,Ba13}, are also discussed to explain the presence of DPNELs in SDSS J1532.

If it is assumed that the dual AGN scenario is responsible for the double-peaked profiles in all narrow lines of SDSS J1532, given the mean peak separation of $\Delta\upsilon \sim$ 340 km/s, as shown in Table 1, then the physical separation between the two putative AGN cores can be simply estimated using a circular motion equation: $\Delta\upsilon=\sqrt{G(M_{1}+M_{2})/r}\times\sin(i)\times\cos(\phi)$, where $M_1$ and $M_2$ denote the total mass of each core, $r$ represents the intrinsic spatial distance between the two cores, and $i$ and $\phi$ are the inclination angle of the rotation plane and the orientation angle, respectively.
Since $i$ and $\phi$ are not observationally constrained, we adopt the maximum projection factor ($\sin(i) \cos(\phi) = 1$; corresponding to~$i = 90^\circ$ and $\phi = 0^\circ$), which provides an upper limit on $r$.

During the dual AGN phase, each black hole is expected to retain the stellar component of its original host galaxy; we therefore use the corresponding total stellar masses $M_{\ast}$ as proxies for $M_1$ and $M_2$.
The total stellar mass can be determined through the $M_{\ast}$-$\sigma_{\ast}$ (the stellar velocity dispersion) relation \citep{Lw13,Za16,Da22}.
Since the spectrum of SDSS J1532 lacks discernible spectroscopic features from the host galaxy, the most practical approach is to utilize the line width of narrow emission lines as a tracer of $\sigma_{\ast}$ \citep{Ne96,Gr05}.
Using the line widths of \oiii~as proxies for $\sigma_{\ast}$, the estimated $M_{\ast}$ values for the two putative cores are $10^{10.72\pm 0.30}\rm M_\odot$ and $10^{10.69\pm 0.29}\rm M_\odot$, respectively, with uncertainties arising from both the line width and the parameters of $M_{\ast}$-$\sigma_{\ast}$ relation. Further details about the calculation of $M_{\ast}$ can be found in \citet{Zh24}.
After treating $M_{\ast}$ as $M_1$ and $M_2$, the upper limit of $r$ is estimated to be 3.14$\pm$2.64 kpc.
The inferred kpc-scale separation is consistent with expectations for dual AGNs, where the two narrow-line regions remain spatially distinct and can produce double-peaked narrow emission lines.
Because the broad-line region emission from the two putative nuclei cannot be disentangled, separate virial black hole mass estimates for each component are not feasible. Therefore, the black hole (BH) mass $M_{\rm BH}$ can be calculated by
the $M_{\rm BH}-\sigma_{\ast}$ relation, which is also applicable in dual cores \citep{Ge00,Jo09,Sa19,Be21}.
Based on the relation in \citet{Ko13} and the widths of \oiii~as tracers of $\sigma_{\ast}$, the values of $M_{\rm BH}$ are $10^{7.61\pm 0.20}\rm M_\odot$ and $10^{7.52\pm 0.20}\rm M_\odot$.

In the rotating disk-like NLR scenario, both of the two sets of emission lines are illuminated by a single ionizing source, and it is expected to be similar physical condition in the narrow line region on both two sides \citep{Xu09}.
Consequently, comparing the line ratios of the two components provides a useful diagnostic for testing the disk interpretation. Although projection effects may introduce differences in the observed line widths and fluxes, large discrepancies in intrinsic line ratios are generally not expected in a simple rotating disk.
Based on the parameters listed in Table 1, the line flux ratios for blue-shifted component and red-shifted component exhibit similar in most narrow emission lines, such as H$\alpha$/[O~{\sc iii}] ratios, broadly consistent with the disk scenario. However, the H$\alpha$/[N~{\sc ii}] ratios show a significant contrast, with values of $5.3 \pm 0.9$ and $2.5 \pm 0.4$ for the blue-shifted and red-shifted components, respectively.
Moreover, according to the criterion in \citet{Sm12}, the similar line flux ratio of blue-shifted component to red-shifted component (0.75$\leq$$F_{b}/F_{r}$$\leq$1.25) often represents rotating disk.
Most of the line flux ratios in double-peaked profiles of SDSS J1532 are consistent with this criterion, but show large difference in [N~{\sc ii}] and [S~{\sc ii}].
Taken together, these results suggest that while some spectral properties are consistent with a rotating disk, the observed line ratio differences indicate that a simple disk model alone may not fully explain the double-peaked structure.

In the AGN-driven outflow scenario, a biconical outflow, moving toward and away from the observer, could naturally generate DPNELs.
Due to to extinction by dust in the galaxy and the outflow itself, the red-shifted components would experience more obscuration, and then typically exhibit smaller line fluxes, compared to the blue-shifted component \citep{Liu10}.
However, in SDSS J1532, the fluxes of the red-shifted and blue-shifted components are comparable within the measurement uncertainties for most narrow emission lines, and no systematic attenuation of the red-shifted component is observed. This behavior is not clearly consistent with the simple expectations of a dust-obscured outflow scenario.
Additionally, a strong outflow is expected to induce velocity stratification between lines originating closer to the core and lines at larger distances \citep{Xu09}.
However, as shown in Table 1, the peak separations $\Delta\upsilon$ between the blue-shifted and red-shifted components are consistent among all these narrow emission lines, showing no evidence for such stratification.
So it disfavors the AGN-driven outflow scenario.

\subsection{QPO properties}

A periodicity of about 220 day is determined in ZTF r-band and g-band light curves, utilizing the LSP method, WWZ method, and epoch-folding method. 
To investigate the origins of the QPO signal observed in SDSS J1532, three models are discussed here: the BBH model, the disk precession model, and the jet precession model.

In the BBH scenario \citep{Gr15,Ko19,Zxg23}, it is proposed to explain long-standing QPO with periodicities spanning from hundreds to thousands of days.
According to the discussion in \citet{Er12}, the space separation of the BBH system can be calculated as 
\begin{equation}
	\begin{split}	
		D_{\rm BBH}\sim 0.432M_8(\frac{P/yr}{2652M_8})^{\frac{2}{3}},
	\end{split}
\end{equation}
where $M_8$ is the $M_{\rm BH}$ in units of $10^8 \rm M\odot$.
After considering the BH mass $M_{\rm BH}$ range from $10^{7.52\pm 0.20}\rm M_\odot$ to $10^{7.61\pm 0.20}\rm M_\odot$ and the periodicity as about 220 day ($P \sim$ 0.6 yr), the space separation between the central BBH might be about 0.95-1.4 mpc.

In the disk precession scenario \citep{Ar13,Zxg23}, the disk precesses relative to its local frame of reference as a result of the rotation of massive black hole.
According to \citet{Er95,St03}, the anticipated period of disk precession can be approximated as $T\sim 1040M_{8}R_{3}^{2.5}\rm yr$, where $R_3$ represents the distance of emission regions to the central BH in units of $10^3$ Schwarzschild radii ($R_G=GM_{\rm BH}/c^2$).
Then the expected $R_3$ is about 0.061-0.095 (61-95~$R_G$).
Additionally, according to the results of \citet{Mo10}, based on microlensing variability analyses of eleven gravitationally lensed quasars, the size of the near-ultraviolet (NUV) 2500~\AA~continuum emission region in SDSS J1532 is given by $\rm log\frac{R_{2500}}{cm}=15.78+0.8\rm log\frac{M_{\rm BH}}{10^9 \rm M\odot}\sim$70-88~$R_G$.
In the framework of the standard thin-disk model \citep{Fr02,Fa16,La23}, the disk temperature decreases with increasing radius. As a result, NUV emission is expected to originate from the hotter inner regions of the accretion disk, while the optical emission is produced at relatively larger radii. However, the radii inferred above for the optical and NUV continua are comparable, indicating that the two bands originate from similar characteristic disk scales rather than from widely separated regions.
Since the ZTF g-band emission originates from regions closer to the central engine, the disk precession model would predict a slightly shorter period in the g-band compared to that in the r-band. However, the periodicities derived from the aforementioned methods are generally consistent between the two bands.
The results do not support the disk precession scenario at the origin of the periodic modulation.

In the jet precession scenario \citep{Ca13,Bh18,Zh21b}, the brightness of a source varies as a result of the helical structure.
However, despite being observed in the Faint Images of the Radio Sky at Twenty-cm \citep{Be95,He15}, SDSS J1532 does not exhibit any apparent radio emissions.
Therefore, jet precession is not a better choice for the optical QPO in SDSS J1532.

\subsection{Future applications}
Although alternative interpretations cannot be ruled out with the current observational evidence, SDSS J1532 remains a good candidate for triple black hole with a close BBH pair.
The double-peaked profiles observed in all narrow emission lines suggest the presence of a dual AGN system on kpc scales, with an estimated projected separation of $\sim$3 kpc. 
Such a separation would correspond to an orbital timescale of several Gyr, far exceeding the temporal coverage of the current photometric data, and therefore cannot be tested through light-curve folding.
In contrast, the QPO likely originates from a much more compact BBH system embedded within one of the nuclei, with an inferred separation around 1 mpc.
Therefore, one of the nuclei in the dual AGN may itself host a close binary black hole. On sub-pc scales, the NLRs are expected to merge into a common region, making split or broadened narrow lines unlikely, whereas the more compact broad-line regions may remain distinct. If both nuclei were type 1 AGNs, double-peaked broad emission lines would be expected. However, the absence of obvious double-peaked broad lines in SDSS J1532 suggests that only one of the compact nuclei likely contributes broad-line emission.
If feasible, future observations with higher spatial or temporal resolution could help confirm or reject these scenarios by providing stronger constraints on the nature of the multiple components.

Evidence for triple black holes not only can validate the hierarchical merger paradigm, but also will significantly advance our understanding of galaxy formation and fundamental physics. 
Furthermore, the presence of triple black holes offers a unique opportunity to study the dynamics of three-body interactions within the framework of general relativity \citep{Me06}. Theoretical predictions indicate that hierarchical systems of closely spaced triple black holes  potentially generate intense bursts of gravitational waves \citep{Iw06, Lo08}. These gravitational waves hold the potential for detection by low-frequency gravitational wave experiments in the future \citep{Am10,Ca18,Ca20,Au23}.

\section{Summary and Conclusions}
Main summary and conclusions are as follows. 
\begin{itemize}
\item SDSS J1532 shows the properties of DPNELs in all narrow emission lines with peak separations about 340 km/s, which might be produced by dual AGN in kpc-scale.
\item The similar line flux ratios of the blue-shifted and red-shifted components in most of the DPNELs are broadly consistent with a disk scenario, while the significantly different H$\alpha$/[N~{\sc ii}] ratios suggest that a simple rotating disk model may not fully explain the observed double-peaked structure.
\item The similarity of line fluxes in blue-shifted and red-shifted components and the similarity of peak separations in different double-peaked features disfavour the AGN-driven outflow model.
\item SDSS J1532 shows an optical quasi-periodic oscillation with a period of about 0.6 yr, corresponding to $\sim$10.5 cycles in the ZTF light curve, while the CSS light curve independently covers $\sim$13.5 cycles.
\item Since the optical emission region estimated by disk precession is similar to NUV emission regions and there is no apparent radio emissions in SDSS J1532, it is opposite to the disk precession and jet precession models.
In the BBH model scenario, the estimated distance will be about 1 mpc between the central two BHs.
\item Although alternative interpretations cannot be ruled out with the current observational evidence, the observed double-peaked narrow emission lines and the optical quasi-periodic oscillation are broadly consistent with a triple black hole scenario, in which the kpc-scale double-peaked features may arise from dual AGN and the optical quasi-periodic modulation may be associated with a sub-pc binary black hole.
If SDSS J1532 is confirmed as a triple black hole system with a close BBH, it would signify a pivotal moment in the evolution of galaxy mergers.

\end{itemize}

\begin{acknowledgements}
We gratefully acknowledge the anonymous referee for giving us constructive comments and suggestions to greatly improve our paper.
We thank Paola Severgnini and Fabio Rigamonti (INAF–Brera Astronomical Observatory) and Cristian Vignali (University of Bologna) for their valuable comments and helpful suggestions that greatly improved this work.
This work is supported by the National Natural Science Foundation of China (Nos. 12273013, 12173020, 12373014). We have made use of the data from SDSS DR16.
The SDSS DR16 website is (\url{http://skyserver.sdss.org/dr16/en/home.aspx}).
The MPFIT website is (\url{http://cow.physics.wisc.edu/~craigm/idl/idl.html}).
The ZTF data can be obtained in the web site (\url{https://irsa.ipac.caltech.edu/frontpage/}).
The CSS web site is (\url{http://nesssi.cacr.caltech.edu/DataRelease/}).
The ATLAS web site is (\url{https://fallingstar-data.com/forcedphot/}).
\end{acknowledgements}

\bibliographystyle{aa}

\end{document}